\documentclass{article}
\usepackage[preprint]{neurips_2025}
\usepackage[utf8]{inputenc}
\usepackage[T1]{fontenc}
\usepackage{hyperref}
\usepackage{url}
\usepackage{booktabs}
\usepackage{amsmath}
\usepackage{amsfonts}
\usepackage{nicefrac}
\usepackage{microtype}
\usepackage{graphicx}
\usepackage[font=small,labelfont=bf,justification=centering,singlelinecheck=on]{caption}
\usepackage{subcaption}
\usepackage{xcolor}
\graphicspath{{../results/}{results/}{figures/}{./}}

\newcommand{\fnote}[1]{\par\vspace{3pt}{\footnotesize\raggedright\textit{Notes.} #1\par}}

\makeatletter\renewcommand{\@notice}{}\makeatother

\title{{\Huge\scshape Doppelganger}\\[0.4em] Sound Effects and Their Synthetic Twins}

\author{%
  Elliott Ash \\
  ETH Z\"urich \\
  \texttt{ashe@ethz.ch} \\
  \texttt{https://github.com/elliottash/doppelganger}
}

\begin{document}
\maketitle

\begin{abstract}
Audio-conditioned generators now produce synthetic sound effects from real recordings, so the real
and synthetic versions of an event increasingly coexist in sound libraries and in the corpora used to
train audio models---yet no benchmark measures whether a representation can match a synthetic clip to
the specific real recording it was generated from. I introduce Doppelganger, a benchmark for
matching sound effects across the synthetic--real boundary, pairing $10{,}420$ real clips across
$34$ everyday sound events each with an audio-conditioned synthetic twin, alongside a controlled
$7$-class corpus. Off-the-shelf audio encoders do not cross the boundary cleanly. Making the
embedding ignore the boundary the standard way---training it on sound-event labels---works on
familiar sounds but backfires on new ones, dropping below the untrained encoder. Training on the
pairs instead---a clip and its own synthetic twin---generalizes. On sound events held out
of training, it recovers the exact real source about $80\%$ of the time (up from $61\%$ untrained; chance $0.03\%$),
whereas no objective meaningfully improves category-level recognition on those unseen events. The learned matching is
specific to one generator---it survives changes to that generator's settings but not a switch to a
different generator, and collapses for the text-only generators tested. A human annotation baseline ($49$
listeners) lands well above chance but below the models on the same trials. Synthetic twins fool people into calling them
real about $29\%$ of the time, yet a generator-specific detector separates these audio-conditioned twins from real recordings perfectly.
\end{abstract}

\begin{figure}[h]
\centering
\caption{The synthetic--real matching task and the instance-versus-category comparison on unseen sound events.}
\label{fig:teaser}
\includegraphics[width=\textwidth]{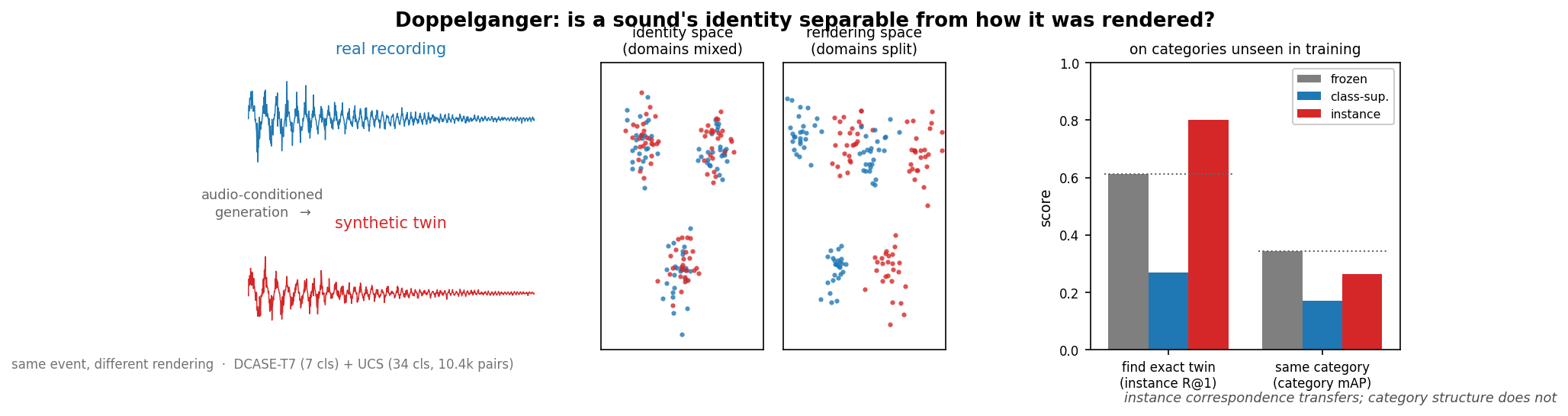}
\fnote{\textbf{Left:} each real recording is paired with an audio-conditioned synthetic twin of the
same event; I ask whether an embedding can represent the event identity while ignoring how it
was rendered. \textbf{Middle:} an invariant embedding mixes the two domains within each event
cluster, while a sensitive one splits them---the two axes this benchmark separates.
\textbf{Right:} on sound types never seen during training, an embedding trained to match each clip
to its own synthetic twin (red) finds the exact real twin far more often than the untrained
baseline (dotted line), while no training meaningfully improves retrieval of other clips of the same kind.
Matching a specific sound to its twin transfers to new sound types; recognizing the broad sound event
does not.}
\end{figure}

\section{Introduction}
Audio-conditioned generators now turn real sound effects into synthetic variants, so the real and
synthetic versions of an event increasingly coexist---side by side in sound libraries, and mixed
together in the web-scale corpora used to train audio models. This raises questions the field's usual
tools do not answer---questions not about whether two clips share a sound event, but about
whether one specific synthetic clip is the counterpart of one specific real recording.
Single-domain embedding benchmarks
\citep{turian2022hear,heigold2026mseb} score whether an embedding groups similar-sounding clips, but
treat all audio as one domain. Distributional generative
metrics such as Fr\'echet Audio Distance~\citep{kilgour2019fad} ask whether a generator's
output distribution resembles real audio, not whether a given output preserved the event it
was conditioned on. And synthetic-audio detection~\citep{ouajdi2024deepfake,yin2025envsdd} asks only
real-or-fake, discarding the correspondence between a synthetic clip and its source.

I argue these are facets of one capability---representing the identity of a sound event across
the synthetic--real boundary (Figure~\ref{fig:teaser})---and pose it as retrieval:
\begin{quote}
Can a representation recognize that a synthetic clip and a real recording are counterparts of
the same event, and does that ability transfer to event types not seen during training?
\end{quote}
Transfer is the objective. No benchmark can enumerate every sound, so a representation is useful only if
the correspondence survives on sound events it never saw. I frame the task as retrieval rather than
detection---detection rewards exploiting the synthetic--real gap, whereas
retrieval rewards removing it, which is what cross-domain search, real/synthetic
deduplication, and clip-by-clip generator evaluation actually need.

This work shows that the correspondence between a clip and its audio-conditioned
synthetic twin is a learnable object that transfers across sound events. Sound-event identity is not, and
class-supervised invariance---the standard recipe---actually degrades unseen sound events below the
untrained baseline. The same machinery yields a complementary sensitive head whose
real-vs-synthetic direction separates a generator's outputs from real audio clip-by-clip: a paired
complement to FAD that audits what a generator did to a specific input rather than to a
distribution. Both capabilities turn out to be generator-specific. The correspondence vanishes for
text-only generators, and the sensitive axis rates an unseen generator's clips as nearly
real---which bounds both claims, and points at how this benchmark can test and guide generators
that condition on audio.

\paragraph{What this is useful for.} The released embeddings support (i) cross-domain
retrieval---find the real source of a synthetic clip, or search a real library with a
generated query; (ii) dataset hygiene---cluster synthetic derivatives with their real sources
to catch leakage and near-duplicate contamination before they enter a training set; and (iii)
generator evaluation---scoring, one clip at a time, whether a synthetic-sound model preserved
the event it was conditioned on, and ranking outputs or systems by it.

\paragraph{Contributions.}
\textbf{(C1)} a two-part benchmark---a controlled 7-class corpus (DCASE-T7, from the Detection and Classification of Acoustic Scenes and Events challenge) and a diverse,
instance-paired 34-event corpus---with leakage-safe splits and a reproducible generation recipe.
\textbf{(C2)} a measured dissociation: an instance-contrastive objective generalizes synthetic--real
instance correspondence to unseen sound events (mean full-gallery R@1 $0.800$, with positive margins in all five folds),
whereas class-supervised invariance does not (and degrades unseen sound events below baseline);
robust over five folds and six encoders---supervised, audio-text, and self-supervised---five of
which did not participate in corpus verification.
\textbf{(C3)} a complementary sensitive head that separates a generator's outputs from real
recordings clip-by-clip (within-generator AUC, area under the ROC curve, $1.0$)---and the finding that it shares the instance
head's boundary: it neither grades fidelity continuously nor transfers to an unseen generator, so
realness detection is per-generator too. \textbf{(C4)} released artifacts---benchmark, recipe,
CLAP-verification, trained heads, and a reusable transform that adjusts any new clip's embedding.

\section{Related Work}
Four lines of work bound this task. Table~\ref{tab:prior} places them along the axes that distinguish
Doppelganger---which domains they span, whether they match at the sound event or the instance level,
where their synthetic audio comes from, and how they evaluate---and I discuss each in turn. The
short version: embedding benchmarks match within a single domain, detection exploits the
synthetic--real gap, and generative metrics score realism distributionally. None pairs each real clip
with an audio-conditioned synthetic twin and measures instance-level correspondence across the
boundary.

\begin{table}[t]
\centering
\caption{Doppelganger compared with related audio benchmarks.}
\label{tab:prior}
\renewcommand{\arraystretch}{1.35}
\resizebox{\textwidth}{!}{%
\begin{tabular}{lcccc}
\toprule
benchmark & domain split & correspondence & synthetic source & evaluation \\
\midrule
HEAR / MSEB / MAEB & single-domain & sound event & --- & class / retrieval / cluster \\
EnvSDD / deepfake-audio & synth vs.\ real & --- (exploit gap) & mixed generators & detection (classifier) \\
FAD & real vs.\ generated & --- & generators & distributional distance \\
\textbf{Doppelganger (ours)} & synth $\leftrightarrow$ real & instance + sound event & audio-conditioned twin & retrieval (R@1, mAP) \\
\bottomrule
\end{tabular}}
\fnote{Rows are prior lines of work; columns are the axes that separate them. domain split:
whether the benchmark spans one audio domain or both real and synthetic. correspondence: the
unit two clips are matched on---sound event, or the specific instance (``exploit gap'' means the method
only separates real from synthetic, matching nothing). synthetic source: where synthetic
audio, if any, comes from. evaluation: the metric. Doppelganger is the only row that pairs
each real recording with an audio-conditioned synthetic twin and evaluates both sound event- and
instance-level correspondence across the synthetic--real boundary.}
\end{table}

\paragraph{Audio-representation benchmarks (row 1).} HEAR~\citep{turian2022hear} standardized
frozen-embedding evaluation across dozens of audio tasks, and the recent massive benchmarks
MSEB~\citep{heigold2026mseb} and MAEB~\citep{elassadi2026maeb} scale this to classification,
retrieval, and clustering over many datasets. They ask whether an embedding groups semantically
similar clips, but treat audio as a single domain. None tests invariance to how a sound was
produced, or asks whether a synthetic clip can be matched to the specific real recording it imitates.
Embedding-based retrieval and representation repair are more developed in text-as-data
settings, from sentence embeddings and dense passage
retrieval~\citep{reimers2019sentencebert,karpukhin2020dpr} to relevance annotation for
retrieval-augmented generation~\citep{ni2025diras} and the removal of document-source confounds from
embedding similarity and clustering~\citep{fan2025medium}. Doppelganger brings this
retrieval-and-representation evaluation perspective to audio. It adds exactly the missing axis and
evaluates the same frozen encoders these benchmarks
rank---CLAP (\citealp{wu2023clap}; \citealp{elizalde2023clap}), PANNs~\citep{kong2020panns},
AST~\citep{gong2021ast}, and the SSL families BEATs~\citep{chen2023beats}, M2D~\citep{niizumi2024m2d},
and AudioMAE~\citep{huang2022audiomae}---under a cross-domain, instance-level protocol.

\paragraph{Synthetic-audio detection (row 2).} The closest prior use of the core corpus is
deepfake-environmental-audio detection: \citet{ouajdi2024deepfake} train a real-vs-synthetic
classifier on the DCASE-T7 Foley set, and EnvSDD~\citep{yin2025envsdd} scales detection to a large
benchmark. These pursue the opposite direction---they learn representations that exploit
the synthetic--real gap, whereas I ask when that gap can be removed while event identity is
kept. I reproduce their objective as a controlled sensitive head and show it is one end of a
single axis whose other end (the invariant head) supports cross-domain retrieval. Detection
and correspondence are thus two readouts of the same representation, not separate problems.

\paragraph{Generative audio and its evaluation (row 3).} The synthetic domain here is produced by
modern generators---Stable Audio Open~\citep{evans2024stableaudioopen} and its predecessor
\citep{evans2024stableaudio}, AudioGen~\citep{kreuk2023audiogen}, and AudioLDM~\citep{liu2023audioldm}.
Their outputs are almost always scored by Fr\'echet Audio Distance~\citep{kilgour2019fad}, which
compares the distribution of generated audio to real audio and therefore cannot say whether a
particular output preserved the event it was conditioned on. That is an open gap for
audio-conditioned generation specifically. The sensitive head gives a paired, clip-level realness
score, and the instance protocol measures whether a generator's output stays identifiable with its
source---diagnostics FAD does not provide.

\paragraph{Domain generalization and contrastive learning (the methods).} I build on
domain-adaptation tools---Proxy-A-distance~\citep{bendavid2010theory}, domain-adversarial
training (DANN)~\citep{ganin2015dann}, Deep CORAL (correlation alignment)~\citep{sun2016coral}, and invariant risk minimization (IRM)~\citep{arjovsky2019irm}---and
on the contrastive family, from supervised contrastive learning~\citep{khosla2020supcon} to instance
discrimination~\citep{wu2018instance} and contrastive pretraining
(\citealp{chen2020simclr}; \citealp{radford2021clip}). The invariant head pursues the same goal
as linear concept erasure---removing one factor (here the real-vs-synthetic rendering domain) from an
embedding while preserving another (event identity), as \citet{fan2025medium} do to deconfound
document embeddings. I find that class-level domain-invariance (supcon plus
adversarial alignment) overfits the training taxonomy and fails on unseen sound events, while an
instance-level objective generalizes---a domain-generalization phenomenon in the synthetic--real
setting that, to my knowledge, has not been reported for audio.

\section{The Doppelganger Benchmark}
\label{sec:bench}
\paragraph{Core corpus (DCASE-T7).} The DCASE-2023 Task-7 corpus~\citep{choi2023foley} provides a ready-made, event-aligned dataset of real and synthetic audio clips spanning 7 classes. It comprises 5{,}550 real clips, sourced from FSD50K~\citep{fonseca2022fsd50k}, UrbanSound8K~\citep{salamon2014urbansound}, BBC, and Freesound. For the experiments, I adopt DCASE's source-disjoint dev/eval split as train-val/test for the real clips, with the synthetic clips hashed into the same train/val/test splits (Appendix~\ref{app:datasheet}). In addition, the corpus includes 25{,}900 synthetic clips generated by 37 different systems. This setup offers a controlled environment for measuring the clean gap between real and synthetic data, as well as for conducting leave-one-class-out probes. Table~\ref{tab:corpus} panel (a) provides an overview.

\begin{table}
\centering
\caption{Corpus composition: clip counts by split (a) and UCS sound morphologies (b).}

\medskip 

\label{tab:corpus}
\small
\begin{minipage}[t]{0.52\textwidth}\centering
\textbf{(a) clip counts by corpus, domain, split}\\[3pt]
\renewcommand{\arraystretch}{1.25}
\begin{tabular}{llccc}
\toprule
corpus & domain & train & val & test \\
\midrule
DCASE-T7 & real  & 4{,}150 & 700 & 700 \\
DCASE-T7 & synth & 18{,}165 & 3{,}884 & 3{,}851 \\
UCS & real  & 6{,}636 & 719 & 3{,}065 \\
UCS & twins & 6{,}636 & 719 & 3{,}065 \\
\bottomrule
\end{tabular}
\end{minipage}\hfill
\begin{minipage}[t]{0.46\textwidth}\centering
\textbf{(b) UCS composition by morphology}\\[3pt]
\renewcommand{\arraystretch}{1.15}
\begin{tabular}{lccc}
\toprule
morphology & events & clips & med.\ len (s) \\
\midrule
transient  & 11 & 3{,}627 & 3.0 \\
texture    &  7 & 2{,}095 & 10.4 \\
tonal      &  4 & 1{,}416 & 5.3 \\
mechanical &  5 & 1{,}317 & 11.8 \\
vocal      &  4 & 1{,}141 & 4.7 \\
ambience   &  2 &   473  & 12.1 \\
whoosh     &  1 &   351  & 0.8 \\
\midrule
total      & 34 & 10{,}420 & 5.7 \\
\bottomrule
\end{tabular}
\end{minipage}
\fnote{\textbf{(a)} Clip counts by corpus, domain, and split. DCASE-T7 uses DCASE's source-disjoint
eval set as the test split; UCS uses FSD50K's train/val/eval splits, keyed on the Freesound uploader
so crops of one recording never straddle a boundary. Counts are the $10{,}420$ CLAP-verified real
clips retained from $13{,}579$ candidates ($77\%$), each paired with one audio-conditioned synthetic
twin, so the real and twin rows match. Instance retrieval queries a synthetic twin against the real
gallery, so the $3{,}065$ real test clips form the retrieval gallery used throughout. \textbf{(b)} The
34 UCS sound events grouped into the seven broad morphologies (families of how a sound is physically
produced) they span. Per sound event, the corpus holds $93$--$410$ real clips (median $338$). med.\ len is the median over a
morphology's events of each event's median clip duration (total = corpus-wide median).
Real-clip length (natural, before the fixed $5$\,s analysis window) runs from $0.3$\,s to the FSD50K
$30$\,s cap, median $5.7$\,s and right-skewed (mean $8.6$\,s): $46\%$ of clips are under $5$\,s and
$33\%$ exceed $10$\,s. Whoosh and transient events are the shortest; mechanical, texture, and
ambience the longest.}
\end{table}

\paragraph{UCS corpus.} Evaluating generalization requires a dataset with many sound events and instance-level pairs. I annotate real audio using the Universal Category System (UCS), mapping the 200 AudioSet~\citep{gemmeke2017audioset} labels from FSD50K~\citep{fonseca2022fsd50k} onto 34 UCS sound events that cover all sound morphologies: transient, texture, tonal, whoosh, vocal, mechanical, and ambience. I verify every clip using zero-shot CLAP~\citep{wu2023clap}, retaining a clip only if its label ranks in the top-5 by audio--text cosine similarity, which keeps 10{,}420 out of 13{,}579 clips ($77\%$), effectively removing FSD50K's noisiest multi-labels. For each verified anchor, I generate a Stable-Audio-Open audio-init twin~\citep{evans2024stableaudioopen} conditioned on the real clip, resulting in 10{,}420 instance pairs. The dataset is split following FSD50K's train/val/eval partitions, keyed by Freesound uploader to ensure that crops from the same recording never cross split boundaries. See Table~\ref{tab:corpus} for summary statistics.

\paragraph{Tasks and metrics.} The two tasks differ in what counts as a correct match. In
sound-event retrieval, I query with a clip and rank clips from the other domain. A match is
correct if it shares the query's sound event. I summarize this with mean average precision (mAP)---how
highly the same-event clips rank, from $0$ to a perfect $1$---and report it against a same-domain
control, so the headline domain gap is control minus cross-domain mAP---the retrieval performance lost
purely by crossing the synthetic--real boundary. In instance retrieval (UCS), only the query's own
twin counts as correct (its exact cross-domain counterpart, sharing an instance id). I report R@1,
the fraction of queries whose top-ranked match is the true twin, and mean reciprocal rank (MRR), which
also gives partial credit when the twin ranks second or third. The headline instance-retrieval numbers
carry bootstrap confidence intervals, and on the UCS corpus I use a $k$-fold leave-classes-out protocol---train with some
sound events withheld, test only on those (Sec.~\ref{sec:diss}). Full per-event and per-split
statistics and the datasheet are in the appendix.

\section{Method}
\label{sec:method}
The pipeline consists of four steps: employing a fixed backbone, attaching a head whose objective determines the preserved geometry, training with pair-aware batches, and using an evaluation protocol specifically designed to prevent the headline number from being inflated by a favorable gallery.

\paragraph{Step 1: Frozen backbone.} I extract embeddings from each frozen encoder---CLAP ($512$-d), PANNs ($2048$-d), and AST ($768$-d). This approach ensures that every comparison isolates the contribution of the head beyond the pretrained representation. Since the backbone's pretraining is uncontrolled, throughout the experiments, ``unseen'' always refers to data not seen by the head.

\paragraph{Step 2: Head and objective.} The head is a compact MLP ($d\!\to\!512\!\to\!256$ with batch-norm and ReLU) with $\ell_2$-normalized output, and the choice of objective is the independent variable. The three contrastive heads share a single loss and differ only in which clips count as positives. For normalized outputs $z_i$ and temperature $\tau$, the supervised contrastive loss over a batch is
\begin{equation*}
\mathcal{L}=\sum_i \frac{-1}{|P(i)|}\sum_{p\in P(i)}\log\frac{\exp(z_i^{\top} z_p/\tau)}{\sum_{a\neq i}\exp(z_i^{\top} z_a/\tau)},
\end{equation*}
where only the positive set $P(i)$ changes between heads. For the invariant head, $P(i)$ is the set of same-event clips, with cross-domain pairs up-weighted (optionally with DANN, CORAL, or IRM). This pulls same-event clips together so that real and synthetic overlap within each event cluster, closing the retrieval gap (though, as shown later, a linear probe can still read the domain). For the sensitive head, $P(i)$ is the same-domain clips within a sound event. It is the deliberate mirror, organizing the space by rendering so that the real-vs-synthetic direction becomes a fidelity axis. For the instance head, $P(i)$ is simply clip $i$'s own generated twin, so it learns the synthetic-to-real mapping rather than class identity. As a compute-matched control, I also train a plain cross-entropy classifier head that matches the instance head in capacity and schedule but is supervised on class labels. This isolates whether any failure arises from the contrastive form or from the use of labels. In the results these appear as the invariant, sensitive, and instance heads, with the class-supervised objectives labelled class-supcon (the invariant head's contrastive objective) and class-CE (the cross-entropy classifier control).

\paragraph{Step 3: Pair-aware training.} Since the instance objective requires both a clip and its twin to be present in the same batch to form a positive pair, I batch by instance id. All heads share the same optimizer, schedule, and capacity, ensuring that any observed differences are attributable solely to the objective: AdamW (learning rate $10^{-3}$, weight decay $10^{-4}$), batch size $1{,}024$, $50$ epochs, and supcon temperature $\tau=0.1$. Exact configurations, seeds, and per-fold class assignments are released with the code.

\paragraph{Step 4: Evaluation protocol.} Sound-event retrieval is scored by mAP against a same-domain control, with the gap defined as control minus cross-domain performance. Instance retrieval is scored by R@1 and MRR against the full real test gallery, which includes all sound events as distractors rather than just a held-out-event pool. The gallery size $N$ and chance level $1/N$ accompany each result. Instance retrieval is directional. The reported direction queries a synthetic clip against a real gallery (synthetic$\to$real), the setting relevant to recovering a real source for a generated sound, and the reverse direction is analogous. Generalization is assessed with a deterministic $k$-fold leave-classes-out protocol (Sec.~\ref{sec:diss}), where each fold trains with certain sound events withheld and evaluates only on those unseen events, reporting bootstrap $95\%$ confidence intervals and per-fold paired margins.

\section{Results: Does sound identity survive in embedding space?}
\label{sec:exp}

\subsection{How well do frozen encoders separate event from rendering?}
\paragraph{The frozen synthetic--real gap.} I investigate whether off-the-shelf encoders exhibit a synthetic--real gap by evaluating frozen CLAP and PANNs~\citep{kong2020panns} on both corpora, comparing retrieval within the same domain to retrieval across domains. The results are presented in Table~\ref{tab:frozen}. In every case, the synthetic--real distinction is present and linearly accessible. For CLAP, cross-domain sound-event retrieval mAP drops by $0.161$ on DCASE-T7 and $0.112$ on UCS compared to the same-domain baseline; for PANNs, the drops are $0.124$ and $0.063$, respectively. A linear domain probe separates real from synthetic at $0.88$ to $0.90$ for CLAP. While absolute mAP is lower on the 34-event UCS corpus than on the 7-event DCASE-T7, the gap and the domain signal remain evident in both. This indicates that sound identity is present but is partially entangled with the method of rendering.

\begin{table}[t]
\centering
\caption{Frozen-encoder cross-domain retrieval gap and domain separability (both corpora, test split).}
\label{tab:frozen}
\renewcommand{\arraystretch}{1.2}
\begin{tabular}{llccccc}
\toprule
encoder & corpus & control & synth$\to$real mAP & gap & PAD & domain-probe \\
\midrule
CLAP        & DCASE-T7 & 0.935 & 0.774 & $+0.161$ & 1.27 & 0.90 \\
CLAP        & UCS      & 0.456 & 0.343 & $+0.112$ & 1.51 & 0.88 \\
PANNs CNN14 & DCASE-T7 & 0.798 & 0.674 & $+0.124$ & 1.39 & 0.90 \\
PANNs CNN14 & UCS      & 0.318 & 0.255 & $+0.063$ & 0.88 & 0.72 \\
\bottomrule
\end{tabular}
\fnote{Frozen encoders on each corpus's test split. control = same-domain sound-event mAP (real
query, real gallery, with each query's own clip excluded from its gallery); synth$\to$real mAP =
cross-domain sound-event retrieval (synthetic query, real gallery); gap = control $-$ cross, the mAP lost by crossing the domain boundary; PAD =
Proxy-A-distance (higher = domains more linearly separable); domain-probe = accuracy of a linear
real-vs-synthetic classifier. Absolute mAP is lower on UCS because it has 34 sound events versus
DCASE-T7's 7, but the gap and the domain probe persist on both corpora: crossing the synthetic--real
boundary costs retrieval accuracy, and a linear probe separates the domains well above chance
everywhere (CLAP $0.88$--$0.90$). The gap is genuine and linearly accessible, not specific to one
corpus.}
\end{table}

\subsection{What do the supervised heads learn---and where do they break?}
The class-supervised invariant head reveals a two-part pattern. It performs well within the training taxonomy but fails as soon as a sound event is held out. These are the first two steps in the paper's argument. The third, discussed in Sec.~\ref{sec:diss}, is that instance-pair training addresses this failure.

\paragraph{The invariant and sensitive heads.} I train two heads that share the same contrastive framework but flip the label, and evaluate them on each corpus's test set with every event seen in training. This isolates what class-label supervision does to the embedding within the taxonomy. The invariant head brings same-event clips together across the domain boundary, while the sensitive head, its intentional counterpart, separates real from synthetic within each event. Table~\ref{tab:heads} demonstrates that these heads are geometric mirror images on both corpora. The invariant head reduces the by-domain silhouette to $0.00$, resulting in real and synthetic clips overlapping within each event cluster, whereas the sensitive head increases it to $0.75$ on DCASE-T7 and $0.85$ on UCS. On the smaller 7-event DCASE-T7 corpus, the invariant head nearly eliminates the retrieval gap, reducing it from $0.161$ to $0.018$. On the 34-event UCS corpus, the gap narrows but does not close, with cross-domain retrieval improving from $0.343$ to $0.558$. Appendix Fig.~\ref{fig:umap} visually illustrates this reversal, and Appendix Table~\ref{tab:ablation} confirms that plain contrastive supervision achieves this effect, with DANN, CORAL, and IRM providing no additional benefit. This demonstrates invariance to rendering, achieved within the training taxonomy.

\begin{table}[t]
\centering
\caption{Retrieval gap and domain silhouette for the invariant and sensitive heads (within-taxonomy, both corpora).}
\label{tab:heads}
\renewcommand{\arraystretch}{1.2}\small
\begin{tabular}{llccccc}
\toprule
head & corpus & control & cross & gap & sil$_{\text{ev}}$ & sil$_{\text{dom}}$ \\
\midrule
frozen    & DCASE-T7 & 0.935 & 0.774 & 0.161 & 0.18 & 0.04 \\
frozen    & UCS      & 0.456 & 0.343 & 0.112 & 0.07 & 0.05 \\
invariant & DCASE-T7 & 0.988 & 0.970 & 0.018 & 0.75 & 0.00 \\
invariant & UCS      & 0.658 & 0.558 & 0.100 & 0.23 & 0.00 \\
sensitive & DCASE-T7 & 0.206 & 0.157 & 0.049 & $-0.04$ & 0.75 \\
sensitive & UCS      & 0.129 & 0.040 & 0.089 & $-0.21$ & 0.85 \\
\bottomrule
\end{tabular}
\fnote{Each head evaluated on both corpora's test splits (within-taxonomy: every event seen in
training). control = same-domain sound-event mAP (real query, real gallery, each query's own clip excluded); cross =
synthetic$\to$real sound-event mAP; gap = control $-$ cross; sil$_{\text{ev}}$/sil$_{\text{dom}}$ =
silhouette score by event class / by domain (higher = tighter clusters on that factor). The two heads
are mirror images on both corpora: the invariant head drives the domain silhouette to $0.00$ (real
and synthetic overlap) while the sensitive head raises it to $0.75$--$0.85$ (real and synthetic split
apart). The invariant head nearly closes the retrieval gap on the small 7-event DCASE-T7 corpus
($0.161$ to $0.018$) and shrinks it on the harder 34-event UCS corpus while lifting cross-domain
retrieval from $0.343$ to $0.558$. Appendix Table~\ref{tab:ablation} shows that plain contrastive
supervision does this work, with DANN, CORAL, and IRM adding nothing.}
\end{table}

\paragraph{Leave-one-class-out.} Does that within-taxonomy success generalize? I hold a single sound event out of training and evaluate the invariant head on it (a leave-one-out probe, distinct from the grouped five-fold leave-classes-out used for the main dissociation), reported in Appendix Table~\ref{tab:closed}. The held-out event performs worse than the untrained frozen baseline (keyboard drops from $0.69$ to $0.43$, gunshot from $0.81$ to $0.31$, rain from $0.75$ to $0.30$), and including more events does not help, since the mean across all 34 UCS events falls from $0.40$ to $0.19$. Thus, class-label supervision not only fails to generalize to unseen events but actually impairs performance.

\subsection{Does instance correspondence generalize across sound events?}
\label{sec:diss}
\begin{figure}[t]
\centering
\caption{Instance, class, and frozen R@1 across six encoders (unseen sound events, full gallery).}
\label{fig:diss}
\includegraphics[width=\textwidth]{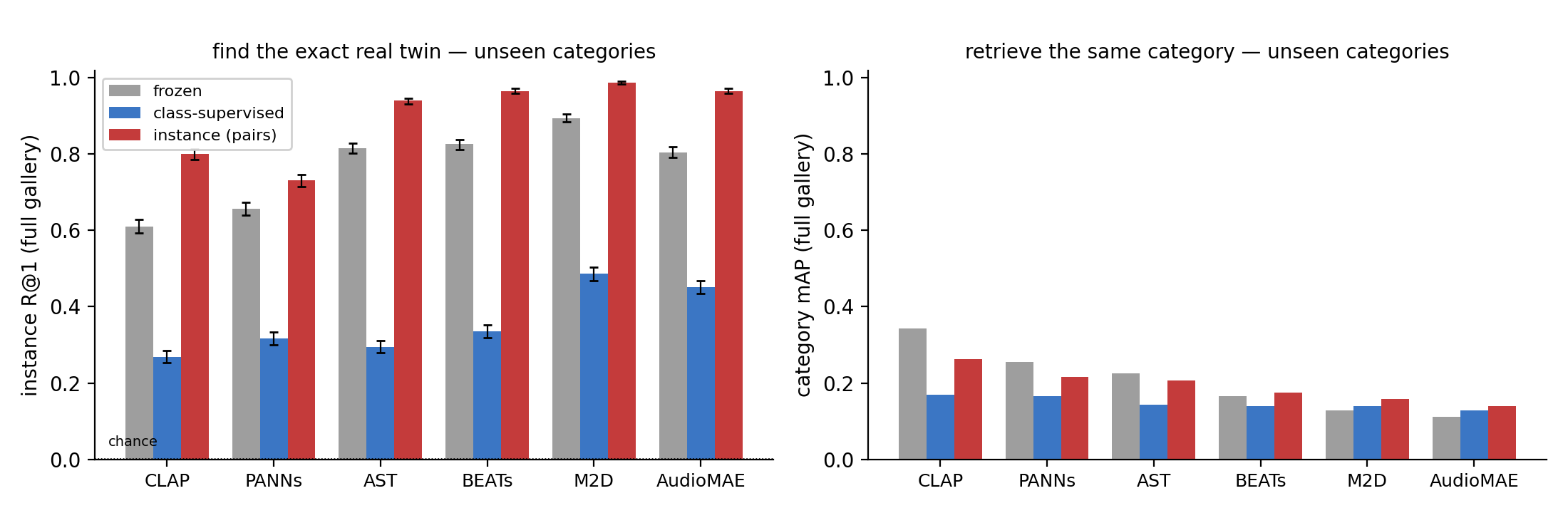}
\fnote{Held-out (unseen) sound events, across six frozen encoders; bars are full-gallery scores
($N=3{,}065$), error bars 95\% bootstrap CIs. \textbf{Left:} synthetic$\to$real instance
retrieval---the instance objective (red) lifts R@1 far above frozen (grey) on every encoder,
while class supervision (blue) falls far below it. \textbf{Right:} sound event
retrieval---low for every objective on every encoder; on the three SSL encoders the heads recover a
sliver of their weak frozen sound-event structure, never approaching frozen CLAP. Instance matching
generalizes across sound events, while category-level retrieval is not improved by any objective.}
\end{figure}

\paragraph{Instance vs.\ class supervision on unseen events.} The central test is whether the correspondence generalizes to sound events not seen during training. I compare three objectives---frozen, class-supervised, and instance-pair---under five-fold leave-classes-out, with results in Table~\ref{tab:diss} and Fig.~\ref{fig:diss}. (Here, ``unseen'' refers to the head, as the frozen backbone's pretraining is uncontrolled.) Against the full real test gallery (a gallery of $N\!=\!3{,}065$ real clips, so one true match and $3{,}064$ distractors from all sound events per query, chance R@1 $=0.0003$), the instance head retrieves the exact real twin with R@1 $0.800$ ($[0.786,0.813]$), outperforming the frozen encoder ($0.611$) in every fold (minimum margin $+0.148$) and class supervision ($0.269$, which is actually detrimental) in every fold (minimum margin $+0.493$). No objective improves sound-event retrieval on unseen events above the frozen encoder (event-mAP near $0.34$). A compute-matched cross-entropy classifier matches class-supcon ($0.282$), indicating that the failure is due to class label training, not the contrastive approach. Only instance-pair supervision generalizes.

\paragraph{Robustness across encoders.} The dissociation could be an artifact of a single encoder. I repeat the leave-classes-out comparison on six frozen encoders from three pretraining families, shown in Table~\ref{tab:encoders}: supervised (PANNs, AST~\citep{gong2021ast}), audio-text contrastive (CLAP), and self-supervised (BEATs~\citep{chen2023beats}, M2D~\citep{niizumi2024m2d}, AudioMAE~\citep{huang2022audiomae}). For all six encoders, the instance head outperforms frozen in every fold, with R@1 ranging from $0.73$ to $0.99$, while class supervision falls well below, from $0.27$ to $0.49$. Five of the six encoders were not involved in the corpus's CLAP verification, so the dissociation is not a result of the CLAP filter. Sound-event retrieval on unseen events remains low throughout. One important detail is that, for the three self-supervised encoders, whose frozen sound-event structure is initially much weaker (full-gallery mAP $0.11$ to $0.17$ compared to $0.34$ for CLAP), the heads recover a small amount of structure (a gain of $0.01$ to $0.05$ event-mAP; per-encoder values in the released results), but still fall far short of frozen CLAP, so the qualitative conclusion is unchanged. The instance mapping transfers almost perfectly, while sound-event structure barely transfers. For the strongest encoders, the frozen twin match is already high---M2D reaches $0.90$ untrained---making the collapse under class supervision the more informative aspect of the dissociation. Per-event instance R@1 and unseen-event embedding geometry are provided in Appendix Table~\ref{tab:percat} and Fig.~\ref{fig:appendix}.

\begin{table}[t]
\centering
\caption{Instance, class, and frozen retrieval across six encoders and three pretraining families (unseen sound events).}
\label{tab:encoders}
\renewcommand{\arraystretch}{1.25}\small
\begin{tabular}{llccc}
\toprule
encoder & pretraining & frozen & class head & instance head \\
\midrule
CLAP        & audio-text contrastive & 0.611 [0.594,0.629] & 0.269 & \textbf{0.800} [0.786,0.813] \\
PANNs CNN14 & supervised (AudioSet)  & 0.657 [0.640,0.673] & 0.316 & \textbf{0.731} [0.715,0.746] \\
AST         & supervised (AudioSet)  & 0.815 [0.802,0.828] & 0.295 & \textbf{0.940} [0.931,0.947] \\
BEATs       & self-supervised (AS2M) & 0.826 [0.812,0.839] & 0.335 & \textbf{0.966} [0.960,0.972] \\
M2D         & self-supervised (AS2M) & 0.895 [0.885,0.906] & 0.486 & \textbf{0.988} [0.984,0.992] \\
AudioMAE    & self-supervised (AS2M) & 0.804 [0.791,0.819] & 0.452 & \textbf{0.966} [0.960,0.972] \\
\bottomrule
\end{tabular}
\fnote{Full-gallery instance R@1 on unseen sound events (five-fold leave-classes-out, synthetic
query, all-sound event real test gallery, $N=3{,}065$, chance $0.0003$); brackets are 95\% bootstrap
CIs over queries (class-head CIs are omitted for space and are in the released results). Per encoder,
a class-supervised and an instance-pair head are trained per fold with identical capacity and schedule. The instance head beats frozen in every fold of every
encoder (minimum per-fold margin $+0.05$), and class supervision falls far below frozen on all
six---the dissociation is not specific to one encoder or pretraining style. PANNs, AST, BEATs, M2D,
and AudioMAE did not participate in the corpus's CLAP-verification.}
\end{table}

\begin{table}[t]
\centering
\caption{Instance versus class supervision on unseen sound events (five-fold leave-classes-out, CLAP).}
\label{tab:diss}
\renewcommand{\arraystretch}{1.25}
\begin{tabular}{lcc}
\toprule
variant (unseen sound events) & event-mAP & instance-R@1 (full gallery) \\
\midrule
frozen        & $0.343$ & $0.611\;[0.594,0.629]$ \\
class-supcon  & $0.169$ & $0.269\;[0.254,0.285]$ \\
class-CE (classifier) & -- & $0.282\;[0.266,0.297]$ \\
\textbf{instance} & $0.262$ & $\mathbf{0.800\;[0.786,0.813]}$ \\
\bottomrule
\end{tabular}
\fnote{Five-fold leave-classes-out on the UCS corpus (CLAP encoder): each fold withholds a set of
sound events from training and is evaluated only on those unseen sound events; rows are training
objectives. event-mAP = full-gallery synthetic$\to$real sound-event retrieval (not computed for the
classifier control, shown as --); instance-R@1 = retrieving the exact real twin of a synthetic query
against the full real test gallery (gallery of $N=3{,}065$ real clips: one true match and $3{,}064$
distractors of all sound events, chance $=0.0003$), with 95\% bootstrap CIs over queries. The instance objective (positive $=$ a clip and its own twin) reaches R@1 0.800, far
above frozen; class-supcon and a compute-matched cross-entropy class-CE classifier both
fall below frozen (0.27--0.28), so the failure is class-label supervision, not the
contrastive form. No objective beats frozen on event-mAP---sound-event structure transfers for none.}
\end{table}

\paragraph{Finer taxonomy.} The 34 UCS categories are intentionally broad, which raises a valid concern. Is ``held-out category'' too easy a test when the categories are coarse? Appendix Table~\ref{tab:fine} addresses this by repeating the leave-classes-out protocol at nearly double the granularity, using $65$ sound events defined by each clip's most-specific FSD50K/AudioSet leaf label (for example, doorbell, gunshot, thunder, computer keyboard rather than ``doors'' or ``weather''). These are single-labelled categories with $\ge 50$ clips each, totaling $8{,}699$ clips, and the same audio, twins, and embeddings are used, with only the labels and head training changed. The dissociation remains unchanged. On unseen fine sound events, the instance head achieves R@1 $0.842$ ($[0.826,0.856]$; gallery $N\!=\!2{,}362$, chance $0.04\%$), outperforming frozen ($0.625$) and class supervision ($0.330$, again below frozen) in every fold, while no objective improves sound-event retrieval (event-mAP frozen $0.406$, instance $0.323$). The qualitative pattern---instance supervision generalizes, class supervision falls below frozen, and sound-event structure does not transfer for either---holds even more clearly at finer granularity, though absolute values are not directly comparable to the 34-event results because the fine gallery is smaller. Approximately $65$ is the practical upper limit here, as FSD50K's finer leaf labels are multi-label and long-tailed, leaving too few single-labelled clips per label to hold out an entire label.

\subsection{Operating characteristics: ranking, direction, and deduplication}
\label{sec:operating}
Appendix Table~\ref{tab:operating} details the practical behavior of the instance head as a tool, all under the same full-gallery, unseen-event protocol. The ranking extends beyond just the top-1. The instance head achieves R@5 $0.945$ and MRR $0.865$ (compared to frozen at $0.801$/$0.696$ and class supervision at $0.419$/$0.344$), so when the exact twin is not ranked first, it is almost always among the top few. The task is also symmetric in direction. Querying with the real clip against the full synthetic gallery (real$\to$synth, the ``which generated clip came from this recording?'' direction) yields R@1 $0.736$ versus $0.547$ for frozen, indicating that the correspondence is not an artifact of the direction reported. The dataset-hygiene use case also functions at a fixed operating point. When half the queries have their twin removed from the gallery, requiring the tool to abstain, sweeping a single cosine threshold yields best-F1 $0.624$ (precision $0.645$, recall $0.604$) for the instance head, compared to $0.479$ for frozen.

How much data does the mapping need? This matters for adapting the recipe to a new corpus and generator. I train the instance head on subsampled pair budgets and evaluate on the full real test gallery with all sound events present in training, shown in Appendix Table~\ref{tab:dataeff}. R@1 increases steadily from $0.681$ at $250$ pairs to $0.843$ at $4{,}000$, compared to $0.611$ for frozen and $0.872$ with all $6{,}636$ pairs. A few hundred audio-conditioned twins provide most of the benefit, and around $2{,}000$ pairs already reach the level of the headline unseen-event result, though here no events are held out of training.

\subsection{Is the sensitive axis a usable realness score?}
\label{sec:sensval}
Does the sensitive axis work as a realness score? I project held-out clips onto its real-minus-synthetic direction and measure how well that separates real audio from synthetic, shown in Appendix Table~\ref{tab:sensitive}. It separates the training generator from real audio almost perfectly, with AUC $1.000$ for real versus Stable-Audio twins at every operating point and $0.999$ against the same model's text-only outputs. However, two clear boundaries appear. First, the axis acts as a detector rather than a graded fidelity measure. Within a generator, its score is nearly flat across operating points and uncorrelated with per-clip twin fidelity (Spearman $-0.10$). Second, it is generator-specific, just like the instance head. ElevenLabs clips, which are synthetic but from an unseen generator, score close to real (AUC $0.747$ against real, and higher than Stable-Audio twins on the axis). Thus, both heads tell a consistent story. The correspondence and the detectable rendering signature are properties of a specific generator family, not of ``synthetic audio'' as a general class. In practical terms, the sensitive head serves as a per-generator audit tool, trained on the generator in use, rather than as a universal realness score.

\subsection{A human annotation baseline}
\label{sec:human}
I collected a human annotation baseline to place the model numbers on a human scale. Fifty-two annotators on Prolific ($49$ analyzed after excluding the three who failed both attention checks), each assigned up to $16$ real-vs-twin discrimination trials (selecting the real clip from two) and $24$ twin-retrieval trials plus $2$ catch trials ($771$ discrimination and $1{,}129$ retrieval judgments analyzed after exclusions), all using the same clips as the models (annotators heard the central $3.5$\,s of the $5$\,s model window). Table~\ref{tab:human} shows that humans perform between chance and the models on both tasks.

\paragraph{Real-vs-twin discrimination.} When distinguishing a real recording from its own synthetic twin (chance $50\%$), humans achieve only $71.3\%$ ($[66.5,76.1]$), making errors on nearly three out of ten pairs, while the sensitive head's real-minus-synthetic axis achieves $100\%$ on the same pairs. This quantifies the premise underlying the dataset-hygiene use case. Human curation would accept a substantial fraction of audio-conditioned twins that the model separates perfectly.

\paragraph{Twin retrieval.} In the task of identifying the exact real source of a synthetic clip among six same-event candidates (chance $16.7\%$), humans reach $82.3\%$ ($[78.2,85.8]$), well above chance, indicating that the correspondence is perceptually real and that the benchmark's pairing is recoverable by independent listeners (majority vote $83.9\%$, with $299/300$ trials having $\ge\!2$ raters). However, frozen CLAP already surpasses human performance ($92.0\%$), and the instance head is near the ceiling ($99.0\%$). The task is difficult for humans but straightforward for the appropriate representation, highlighting the gap exposed by the benchmark.

\begin{table}[t]
\centering
\caption{Human and model accuracy on the two annotation tasks.}
\label{tab:human}
\renewcommand{\arraystretch}{1.25}\small
\begin{tabular}{lccc}
\toprule
task & chance & human & model \\
\midrule
tell the real clip from its twin (2 choices) & 50.0 & 71.3 [66.5, 76.1] & \textbf{100.0} \\
find the real source (6 choices)             & 16.7 & 82.3 [78.2, 85.8] & \textbf{92.0} / \textbf{99.0} \\
\bottomrule
\end{tabular}
\fnote{Accuracy (\%) on the identical trials shown to annotators. $N=49$ analyzed Prolific annotators
(3 of 52 excluded for missing both attention checks), $771$ discrimination and $1{,}129$ retrieval
judgments; human 95\% CIs are bootstrapped over annotators. model: for the two-choice task,
the sensitive head's real-vs-synthetic direction picks the clip it scores as more real; for the
six-choice task, frozen CLAP ($92.0$) and the instance head ($99.0$) each pick the nearest of the same six candidates---a restricted six-choice task, not the full-gallery R@1.
Annotators heard the central $3.5$\,s of the $5$\,s clip the models embed, so the comparison is close
but not exactly matched.}
\end{table}

\subsection{Is it just near-copies? Two controls}
\paragraph{Non-learned dedup baselines.} If the twins were merely waveform-level near-duplicates, standard deduplication tools would catch them. I run three non-learned baselines on the same full-gallery task, reported in Appendix Table~\ref{tab:baselines}. These methods perform poorly: Chromaprint audio fingerprinting, the industry-standard duplicate detector, achieves R@1 $0.066$, while log-mel and MFCC spectrogram cosine reach $0.236$ and $0.229$, respectively, compared to $0.611$ for frozen CLAP and $0.800$ for the instance head. The correspondence resides in the event structure captured by learned representations, not in simple waveform overlap.

\begin{figure}[t]
\centering
\caption{Instance and frozen R@1 versus twin fidelity (fidelity sweep).}
\label{fig:fidelity}
\includegraphics[width=0.82\textwidth]{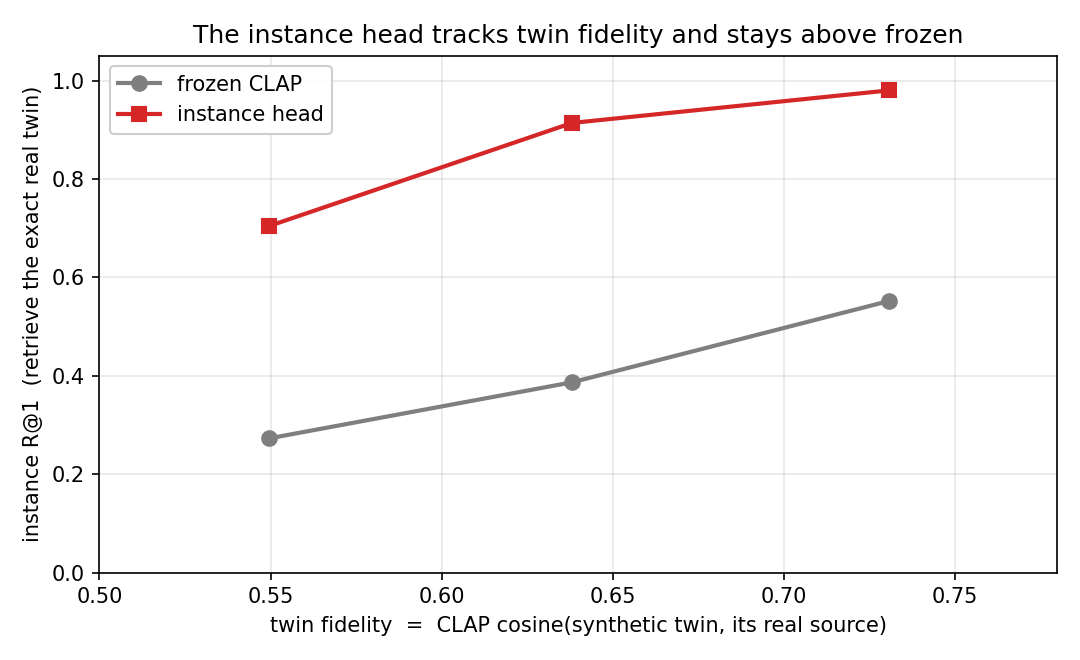}
\fnote{Instance R@1 against measured twin fidelity---the CLAP cosine between each synthetic twin and
its own real source---for frozen CLAP versus the instance head, on in-domain UCS pairs. As twins are
rendered less faithfully (moving left) the instance head degrades gracefully and stays above frozen. The head is
therefore exploiting a learned correspondence, not near-duplicate copies.}
\end{figure}

\paragraph{Fidelity sweep.} Could the instance head simply be matching near-copies? I vary each twin's fidelity to its source and measure retrieval at every level, shown in Fig.~\ref{fig:fidelity}. Across the fidelity range, the instance head maintains a substantial advantage over frozen (a margin of $0.43$ to $0.52$), and performance drops to chance only when the twin is entirely independent of its source. This control is measured in CLAP space and is in-domain; the claim regarding generalization across sound events is supported by the leave-classes-out result above.

\subsection{The generator-specificity boundary: a transfer matrix}
\label{sec:xfer}
Rather than infer the boundary, I measure it directly by generating twin sets under four generation conditions beyond the core corpus: two more Stable-Audio operating points (\texttt{init\_noise} $0.45$ and $0.75$), a second audio-conditioned family (AudioLDM style-transfer at strength $0.5$~\citep{liu2023audioldm}), and a second text-only generator (the Woosh whoosh-domain diffusion model, evaluated on whoosh-sound event anchors). I then cross these with instance heads trained on each family, as shown in Table~\ref{tab:xfer}.

\paragraph{Transfer across operating points and families.} Table~\ref{tab:xfer} shows that heads trained only on Stable-Audio twins at noise $0.6$ transfer effectively to other Stable-Audio operating points, achieving unseen-event R@1 $0.862$ and $0.728$ at noise $0.45$ and $0.75$, with a margin of $0.17$ to $0.22$ over frozen. However, they do not transfer to AudioLDM twins, improving frozen only from $0.061$ to $0.085$, and the reverse transfer is even less effective. A head trained on AudioLDM pairs actually reduces Stable-Audio retrieval, scoring $0.523$ compared to $0.611$ for frozen, even though an AudioLDM-specific mapping can be learned, as the AudioLDM head achieves $0.147$ on its own twins, more than double frozen. Each head thus encodes the rendering map specific to its own generator. One confound acknowledged in the table is that AudioLDM twins have lower fidelity (mean CLAP cosine to source $0.40$, with about $10\%$ degenerate), so family and fidelity are somewhat entangled. The reverse case, where the AudioLDM head lowers Stable-Audio retrieval below frozen, cannot be explained by fidelity alone.

\paragraph{Text-only generators.} Both text-only generators fail in the same way in these auxiliary controls. ElevenLabs and the Woosh generator sit near $2$--$3\%$ R@1, far below the audio-conditioned twins, and no head meaningfully improves them, because the caption bottleneck results in twins that are event-appropriate but not true renderings of the specific clip. The learned mapping is therefore specific to audio-conditioned generation and to the generator family. It serves as an instance-correspondence inverse for a single family, not as a universal synthetic-to-real transformation.

\begin{table}[t]
\centering
\caption{Cross-generator transfer matrix (instance R@1).}
\label{tab:xfer}
\renewcommand{\arraystretch}{1.25}\small
\begin{tabular}{llcccc}
\toprule
twin set & conditioning & fidelity & frozen & SAO-trained head & ALDM-trained head \\
\midrule
SAO, noise 0.45 & audio & 0.78 & 0.696 & \textbf{0.862} & 0.609 \\
SAO, noise 0.6\ (core) & audio & 0.74 & 0.611 & \textbf{0.800} & 0.523 \\
SAO, noise 0.75 & audio & 0.69 & 0.511 & \textbf{0.728} & 0.434 \\
AudioLDM, strength 0.5 & audio & 0.40 & 0.061 & 0.085 & \textbf{0.147} \\
Woosh gen.\ (whoosh events) & text & 0.57 & 0.023 & 0.012 & 0.000 \\
ElevenLabs & text & 0.53 & 0.024 & 0.027 & 0.015 \\
\bottomrule
\end{tabular}
\fnote{Synthetic$\to$real exact-twin retrieval. fidelity = mean CLAP cosine between each twin and its
source. The four audio-conditioned rows use the full real test gallery ($N=3{,}065$, chance $0.0003$);
the two text-only rows are auxiliary and not directly comparable, each with its own query set and
gallery (Woosh: 86 whoosh-event queries; ElevenLabs: 670 train+val queries against an $N=7{,}355$
gallery, as no test-split ElevenLabs twins exist---this auxiliary row is not leakage-controlled, since
it reuses train/val anchors, yet still fails). SAO-trained head = the paper's instance heads
(trained on Stable-Audio noise-0.6 pairs), evaluated under the five-fold unseen-event protocol;
ALDM-trained head = an instance head trained on AudioLDM pairs instead. Rows 1--3: the mapping transfers
across operating points of the training family. Row 4: it does not transfer across families, in either
direction---the ALDM head beats frozen only on its own twins and degrades SAO retrieval below
frozen---though ALDM's lower fidelity partially entangles family with fidelity. Rows 5--6: text-only
generation breaks instance correspondence for every model and no head recovers it.}
\end{table}

\section{Conclusion}
Doppelganger measures whether an audio representation can match a synthetic clip to the specific real
recording it was generated from, and whether that ability survives on sound events unseen in training.
The answer is a dissociation. Instance correspondence generalizes across sound events while no objective
meaningfully improves category-level recognition on unseen ones, and it holds within one generator family but not across families. The broader
lesson is conceptual: what the field calls ``invariance to rendering'' is really two separable
capabilities---recognizing the same specific event across the synthetic--real boundary, and
recognizing the same kind of event---and a representation can acquire one without the other.

The boundaries this exposes map the next questions. Since each generator family carries its own
rendering map, can a head trained on a mixture of families become generator-agnostic, or does adding
families merely average incompatible maps? What survives as the twin is generated from text, image, or
a physical-parameter description rather than from the source audio? And can a representation made
invariant to one rendering process transfer that invariance to another? Each strips away a piece of
the conditioning that makes the present correspondence learnable, moving from matching a sound to its
twin toward representing the event itself, independent of how it was produced.


\clearpage

\bibliographystyle{plainnat}
\bibliography{references}

\clearpage

\appendix
\renewcommand{\thetable}{A\arabic{table}}
\renewcommand{\thefigure}{A\arabic{figure}}
\setcounter{table}{0}
\setcounter{figure}{0}
\section{Datasheet, hosting, and maintenance}
\label{app:datasheet}

\paragraph{Motivation.} The dataset measures whether general-purpose audio embeddings represent the
identity of a sound event across the synthetic--real boundary, operationalized as
cross-domain retrieval at sound event and instance level, with heads that make the representation
deliberately invariant or sensitive to the rendering process. It was assembled by the author. The
core corpus is a re-framing of an existing public corpus (no new audio collection), and the
extensions add generated audio.

\paragraph{Composition (core, DCASE-T7).} $31{,}450$ short ($\le\!4$\,s) mono clips: $5{,}550$ real
and $25{,}900$ synthetic across 7 event classes (dog\_bark, footstep, gunshot, keyboard,
moving\_motor\_vehicle, rain, sneeze\_cough), the synthetic clips coming from 37 generator
identities (9 Track-A and 27 Track-B DCASE-2023 Task-7 challenge systems plus the baseline). Real
audio originates from FSD50K, UrbanSound8K, BBC Sound Effects, and Freesound (provenance columns
\texttt{orig\_dataset}, \texttt{orig\_id} joined from the DCASE metadata). Splits: DCASE
\texttt{eval} is the test set (source-disjoint from \texttt{dev} by construction). \texttt{dev} is
split train/val by original Freesound recording id. Synthetic clips are hashed 70/15/15 with every
generator present in each split. DCASE generators are event-conditional, so the core corpus has
no instance-level pairing.

\paragraph{Composition (UCS corpus).} 34 Universal Category System CatIDs mapped from FSD50K's 200
AudioSet labels (\texttt{src/taxonomy\_ucs.py}), spanning all morphologies. $13{,}579$ real FSD50K
clips selected balanced per CatID, then CLAP-verified (kept iff the clip's audio matches its UCS
label at top-5 by audio--text cosine), retaining $10{,}420$. One Stable-Audio-Open audio-init twin
per verified anchor shares the anchor's \texttt{instance\_id} and split; $670$ ElevenLabs text-only
twins (captioned from each clip's Freesound title and tags) support the cross-generator boundary;
a fidelity-spectrum subset ($1{,}360$ anchors) is generated at multiple \texttt{init\_noise}
operating points plus a text-only condition. Splits follow FSD50K's train/val/eval, keyed on the
Freesound uploader so crops of one recording never straddle a boundary.

\paragraph{Collection and preprocessing.} No primary audio collection; the only new human data are the Prolific annotation choices (Appendix~\ref{app:human}). The pipeline programmatically
wraps released archives---walking the directory layout, joining provenance metadata, assigning
leakage-safe splits, and emitting one manifest row per clip. Clips are loaded mono, resampled,
peak-normalized, and padded or center-cropped to a fixed window so the synthetic--real gap is not
confounded by duration or loudness. Embeddings are cached as aligned \texttt{(ids, emb)} arrays per
encoder. A near-duplicate audit (intra-class cross-split cosine $>0.98$) ships with the code.

\paragraph{Uses.} Intended: benchmarking audio representations for production-invariant retrieval;
training invariance/instance heads; per-generator realness auditing of audio-conditioned
generation; dataset hygiene (real/synthetic deduplication). Not recommended: deploying the
sensitive head as a general ``AI-audio detector''---Sec.~\ref{sec:sensval} shows it is
generator-specific by construction, and it is provided as a scientific control.

\paragraph{Distribution and licensing.} Code is MIT. Real audio is redistributed only where the
source license permits (FSD50K, UrbanSound8K, Freesound CC). The 662 BBC Sound Effects clips are
research-only and are not redistributed---they ship as manifest rows (\texttt{is\_cc=0})
with a fetch script. Synthetic DCASE audio is redistributable under the challenge terms. Generated
twins ship with the full generation log (event, prompt, seed, steps, cfg, noise level) and are
redistributed only when their source clip's license permits derivative outputs; twins of restricted
sources ship as manifest rows with the log instead. Each twin is reproducible from the pinned Stable
Audio Open model version, environment, and logged seed (approximate rather than bit-identical, since
generation is not guaranteed deterministic across hardware and library versions).

\paragraph{Hosting and maintenance.} Code on GitHub (\texttt{github.com/elliottash/doppelganger});
audio, manifests, embeddings, and trained heads on the Hugging Face Hub
(\texttt{huggingface.co/datasets/elliottash/doppelganger}). Manifests are versioned. All derived
artifacts (embeddings, heads, results) regenerate deterministically from the manifest and source
archives, so corrections are shipped as manifest revisions rather than mutated audio.

\section{Human annotation baseline: protocol}
\label{app:human}
Annotators ($N=52$, $49$ analyzed) were recruited on Prolific, desktop with working audio
required, paid \pounds$2.62$ for a task of about $12$ minutes (well above the platform minimum).
The interface presented two blocks: $16$ two-alternative forced-choice trials (a real clip and its
own synthetic twin in random A/B order, ``which is the real recording?'') followed by $24$
six-way retrieval trials (a synthetic query and six same-event real clips, ``which one was this
generated from?'') interleaved with $2$ catch trials whose ``synthetic'' query was in fact a real
recording. All stimuli were the central $3.5$\,s of the same clips the models embed, loudness-
normalized and served under opaque filenames so the answer could not leak through the browser.
An annotator who missed both catch trials was excluded from analysis (3 of 52). Everyone who
finished was paid regardless. Human 95\% CIs are bootstrapped over annotators. Collection stored
only the Prolific ID (used solely for compensation and removed from the released, anonymized data),
per-trial choices, and response times. Annotators consented on the first
screen. Model accuracy in Table~\ref{tab:human} is computed on the identical trial sets.

\section{Within-taxonomy geometry (DCASE-T7)}
\begin{figure}[h]
\centering
\caption{Embedding geometry of the frozen encoder and the invariant and sensitive heads (DCASE-T7).}
\label{fig:umap}
\includegraphics[width=0.92\textwidth]{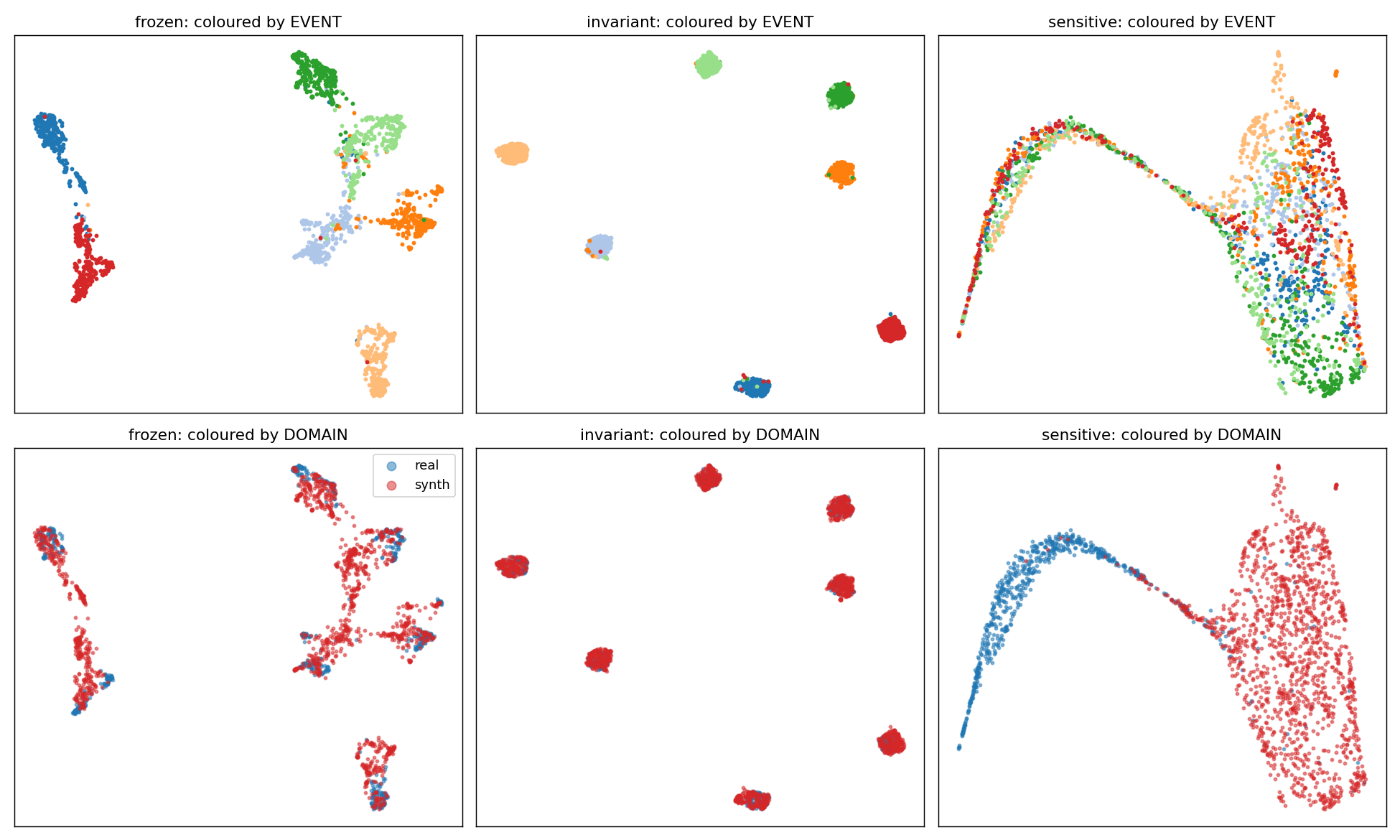}
\fnote{UMAP (Uniform Manifold Approximation and Projection) of the DCASE-T7 test set; columns are the frozen encoder and the two heads. Top row is
coloured by event class, bottom row by domain (real vs.\ synthetic). The invariant head (middle)
tightens event clusters while real and synthetic overlap within each---event identity kept, domain
removed. The sensitive head (right) does the opposite: it splits real from synthetic and smears
the event clusters---domain kept, event removed. Frozen (left) sits between the two. This is the
visual form of the silhouette flip in Table~\ref{tab:heads}.}
\end{figure}

\section{Per-event analysis}
\begin{table}[h]
\centering
\caption{Per-event frozen-CLAP cross-domain retrieval (DCASE-T7).}
\label{tab:perevent}
\renewcommand{\arraystretch}{1.25}\small
\begin{tabular}{lccccccc}
\toprule
& motor & gunshot & dog\_bark & sneeze & rain & footstep & keyboard \\
\midrule
mAP & 0.852 & 0.810 & 0.809 & 0.793 & 0.752 & 0.718 & 0.692 \\
\bottomrule
\end{tabular}
\fnote{Synthetic$\to$real sound event mAP for the frozen CLAP encoder, broken out by DCASE-T7 event
class (sorted high to low). Sustained or broadband events (motor, rain) cross the synthetic--real
boundary best; fast, granular transient and mechanical events (footstep, keyboard) transfer
worst---these are the hardest events to render faithfully and the hardest to retrieve across the gap.}
\end{table}

\begin{table}[h]
\centering
\caption{Per-sound event instance R@1 on unseen sound events (UCS, CLAP, full gallery).}
\label{tab:percat}
\renewcommand{\arraystretch}{1.15}\small
\begin{tabular}{lccc@{\hskip 2em}lccc}
\toprule
cat & $n$ & frozen & instance & cat & $n$ & frozen & instance \\
\midrule
WOOD & 62 & 0.79 & 0.95 & WHSH & 86 & 0.79 & 0.83 \\
BODY & 80 & 0.86 & 0.95 & ANML & 125 & 0.65 & 0.82 \\
DSHS & 121 & 0.81 & 0.93 & BIRD & 81 & 0.52 & 0.81 \\
VOX & 121 & 0.80 & 0.93 & WIND & 26 & 0.42 & 0.81 \\
CREA & 30 & 0.70 & 0.93 & CLTH & 101 & 0.63 & 0.80 \\
CRSH & 33 & 0.85 & 0.91 & BELL & 131 & 0.64 & 0.78 \\
MUSC & 70 & 0.83 & 0.89 & AMB & 17 & 0.12 & 0.76 \\
ALRM & 95 & 0.83 & 0.88 & CRWD & 98 & 0.48 & 0.76 \\
PHON & 120 & 0.82 & 0.88 & CLCK & 85 & 0.60 & 0.75 \\
FOOT & 116 & 0.70 & 0.88 & TOOL & 57 & 0.56 & 0.72 \\
KEYS & 132 & 0.76 & 0.88 & COMP & 120 & 0.53 & 0.71 \\
EXPL & 73 & 0.58 & 0.88 & MECH & 62 & 0.44 & 0.69 \\
GLAS & 129 & 0.71 & 0.88 & ELEC & 40 & 0.45 & 0.68 \\
IMPT & 69 & 0.78 & 0.86 & RAIN & 99 & 0.19 & 0.67 \\
GUN & 109 & 0.58 & 0.85 & VEH & 78 & 0.21 & 0.50 \\
DOOR & 141 & 0.65 & 0.85 & WATR & 121 & 0.31 & 0.50 \\
FIRE & 134 & 0.60 & 0.84 & WTHR & 103 & 0.10 & 0.44 \\
\bottomrule
\end{tabular}
\fnote{Exact-twin retrieval (synthetic query, full real test gallery, $N=3{,}065$) on sound events
held out of head training, pooled over the five folds; UCS CatIDs (four-letter Universal Category
System codes, e.g.\ WOOD = wood, DSHS = dishes, WHSH = whoosh; full mapping in the released
\texttt{src/taxonomy\_ucs.py}) sorted by instance-head R@1 (left column high, right column low).
$n$ = number of queries per sound event. The instance head
improves every sound event over frozen CLAP. The pattern mirrors the DCASE per-event analysis
(Table~\ref{tab:perevent}): discrete, source-specific events (wood, body, dishes, voice) retrieve
best, while sustained textural sound events (weather, water, vehicles, rain)---where distinct
recordings genuinely sound alike---remain hardest.}
\end{table}

\begin{figure}[h]
\centering
\caption{Probe separability and the instance head's unseen-event geometry.}
\label{fig:appendix}
\begin{subfigure}{0.49\textwidth}\includegraphics[width=\linewidth]{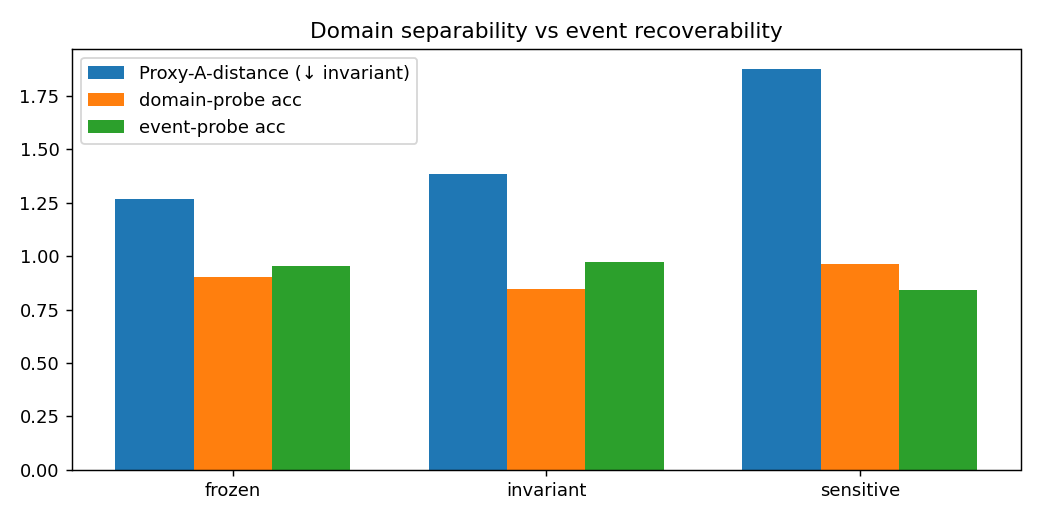}\caption{}\end{subfigure}\hfill
\begin{subfigure}{0.49\textwidth}\includegraphics[width=\linewidth]{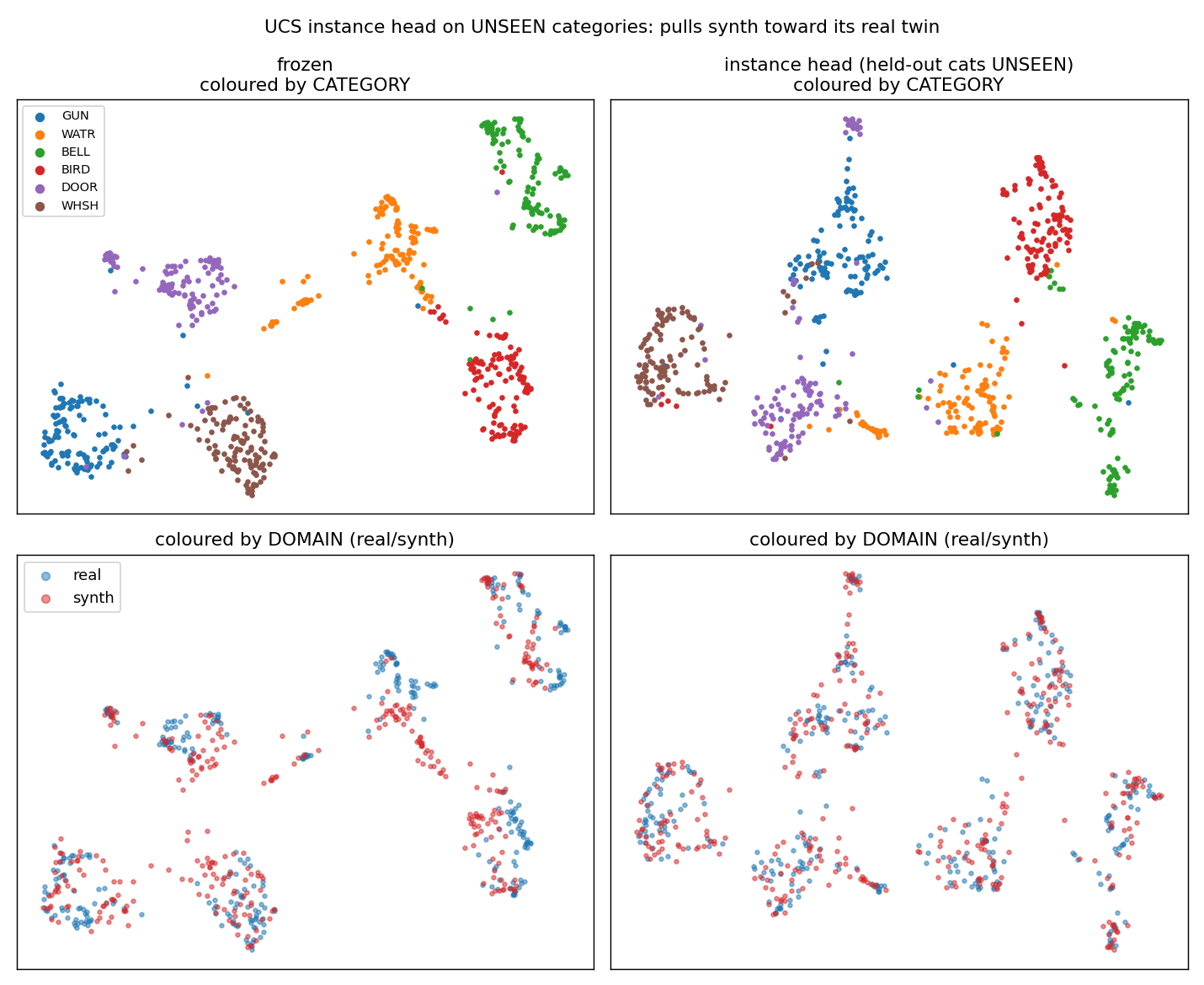}\caption{}\end{subfigure}
\fnote{(a) Proxy-A-distance (PAD) and event/domain linear-probe accuracy across the DCASE-T7 heads.
The sensitive head sharply raises domain separability (PAD $1.88$, domain probe $0.97$). The invariant
head lowers the domain probe modestly ($0.90$ to $0.84$) and keeps event-probe accuracy high ($0.97$),
but does not erase linear domain information---its PAD ($1.39$) stays above frozen's ($1.27$). The
invariant head's invariance shows up as the closed retrieval gap and near-zero domain silhouette
rather than as domain-probe erasure. (b) UMAP of the instance head on sound events unseen
during its training, coloured by domain: real and synthetic mix more thoroughly than under the frozen
encoder, consistent with the head having learned a cross-domain instance correspondence rather than a
real-vs-synthetic split.}
\end{figure}

\section{Supporting result tables}

\begin{table}[h]
\centering
\caption{Bridging objectives compared: plain contrastive versus DANN, CORAL, and IRM (DCASE-T7).}
\label{tab:ablation}
\renewcommand{\arraystretch}{1.25}\small
\begin{tabular}{lcc}
\toprule
invariant-head variant & gap & PAD \\
\midrule
supcon only  & $+0.019$ & 1.38 \\
\;+CORAL      & $+0.017$ & 1.38 \\
\;+DANN       & $+0.019$ & 1.38 \\
\;+DANN+IRM   & $+0.018$ & 1.38 \\
\bottomrule
\end{tabular}
\fnote{Starting from cross-domain supervised contrast (supcon) on the invariant head, adding CORAL,
DANN, or IRM changes the residual gap (control $-$ cross) by less than $0.002$ and the
Proxy-A-distance (PAD, domain separability) not at all. The contrastive term does the work; the
domain-adaptation add-ons are inert here.}
\end{table}

\begin{table}[h]
\centering
\caption{Leave-one-class-out retrieval under class supervision versus frozen (DCASE-T7 / UCS).}
\label{tab:closed}
\renewcommand{\arraystretch}{1.25}\small
\begin{tabular}{lcc}
\toprule
held-out event & frozen & class-supervised (held out) \\
\midrule
keyboard & 0.69 & 0.43 \\
gunshot  & 0.81 & 0.31 \\
rain     & 0.75 & 0.30 \\
\midrule
all 34 UCS events (mean) & 0.40 & 0.19 \\
\bottomrule
\end{tabular}
\fnote{Cross-domain sound-event mAP on a category the class-supervised head never saw in training.
Rows 1--3 are single-event leave-one-out on DCASE-T7; the last row is the mean over the 34 UCS events
under leave-classes-out. In every case the class-supervised head scores \emph{below} the untrained
frozen encoder on the unseen event---the closed-world failure that the instance objective later
repairs.}
\end{table}

\begin{table}[h]
\centering
\caption{Instance versus class supervision at two taxonomy granularities (65 vs 34 sound events, CLAP, unseen events).}
\label{tab:fine}
\renewcommand{\arraystretch}{1.25}\small
\begin{tabular}{lcccc}
\toprule
objective & \multicolumn{2}{c}{65 fine events ($N=2{,}362$)} & \multicolumn{2}{c}{34 events ($N=3{,}065$)} \\
 & event-mAP & instance-R@1 & event-mAP & instance-R@1 \\
\midrule
frozen         & 0.406 & 0.625 [0.606,0.645] & 0.343 & 0.611 \\
class-supcon   & 0.273 & 0.330 [0.312,0.349] & 0.169 & 0.269 \\
instance       & 0.323 & \textbf{0.842} [0.826,0.856] & 0.262 & \textbf{0.800} \\
\bottomrule
\end{tabular}
\fnote{Five-fold leave-classes-out at two granularities: the 34 UCS events (right) and 65 finer events
defined by each clip's most-specific FSD50K/AudioSet leaf label (left; single-labelled, $\ge 50$ clips
each). Same protocol, audio, twins, and embeddings; only the labels and heads differ. The instance
head beats frozen and class-supervision on every fold at both granularities, and no objective improves
sound-event retrieval (event-mAP)---the dissociation is not an artifact of coarse categorization.
Absolute levels differ because the fine gallery is smaller.}
\end{table}

\begin{table}[h]
\centering
\caption{Instance-head ranking depth, retrieval direction, and abstention (UCS, unseen events, full gallery).}
\label{tab:operating}
\renewcommand{\arraystretch}{1.2}\small
\begin{tabular}{llccc}
\toprule
objective & direction & R@1 & R@5 & MRR \\
\midrule
frozen   & synth$\to$real & 0.611 & 0.801 & 0.696 \\
frozen   & real$\to$synth & 0.547 & 0.749 & 0.642 \\
class-supcon & synth$\to$real & 0.269 & 0.419 & 0.344 \\
instance & synth$\to$real & \textbf{0.800} & \textbf{0.945} & \textbf{0.865} \\
instance & real$\to$synth & 0.736 & 0.916 & 0.816 \\
\bottomrule
\end{tabular}
\fnote{Full-gallery retrieval on unseen events ($N=3{,}065$, chance R@1 $=0.0003$). R@5 = fraction of
queries whose true twin is in the top five; MRR = mean reciprocal rank (partial credit for near
misses). The instance head's ranking is deep (R@5 $0.945$) and works in both directions: querying a
real clip against the synthetic gallery (real$\to$synth) gives R@1 $0.736$. A dataset-hygiene
operating point---half the queries have their twin removed, so the tool must also decline---gives
best-F1 $0.624$ (precision $0.645$, recall $0.604$) for the instance head versus $0.479$ for frozen.}
\end{table}

\begin{table}[h]
\centering
\caption{Instance R@1 versus number of training pairs (subsampled, instance head).}
\label{tab:dataeff}
\renewcommand{\arraystretch}{1.2}\small
\begin{tabular}{lccccccc}
\toprule
training pairs & 0 (frozen) & 250 & 500 & 1{,}000 & 2{,}000 & 4{,}000 & 6{,}636 (all) \\
\midrule
instance R@1 & 0.611 & 0.681 & 0.722 & 0.771 & 0.802 & 0.843 & 0.872 \\
\bottomrule
\end{tabular}
\fnote{Instance head trained on random subsamples of the audio-conditioned pairs, with all sound events
present in both training and test (distinct from the leave-classes-out headline protocol), full gallery
$N=3{,}065$; 95\% bootstrap CIs are $\pm 0.015$ throughout. A few hundred pairs already buy most of the
effect, and ${\sim}2{,}000$ pairs reach the headline unseen-event level---practical for adapting the
recipe to a new corpus and generator.}
\end{table}

\begin{table}[h]
\centering
\caption{Sensitive-axis separability across generators and operating points.}
\label{tab:sensitive}
\renewcommand{\arraystretch}{1.2}\small
\begin{tabular}{lcc}
\toprule
clip source & axis score (higher = realer) & AUC vs.\ real \\
\midrule
real recordings         & $+0.63$ & --- \\
Stable-Audio (noise 0.6) & $-0.63$ & 1.000 \\
Stable-Audio (noise 0.9) & $-0.65$ & 1.000 \\
Stable-Audio (noise 1.2) & $-0.64$ & 1.000 \\
Stable-Audio (text-only) & $-0.56$ & 0.999 \\
ElevenLabs (unseen generator) & $+0.41$ & 0.747 \\
\bottomrule
\end{tabular}
\fnote{Held-out clips projected onto the sensitive head's real-minus-synthetic axis (trained on the
UCS train split). AUC uses real as the positive class, so $1.0$ means real and synthetic are perfectly
separated and $0.5$ means indistinguishable. The axis separates real audio from the training generator
essentially perfectly (AUC $1.000$) at every operating point, but two limits show it is a \emph{detector}, not a fidelity
meter: its score is nearly flat across the generator's operating points and uncorrelated with per-clip
twin fidelity (Spearman $-0.10$), and it rates clips from an \emph{unseen} generator (ElevenLabs) as
almost real (AUC $0.747$ against real) and is scored realer than the training generator's own twins
(AUC $0.993$ with ElevenLabs treated as the realer class). Realness detection is thus per-generator, mirroring the instance
head's boundary.}
\end{table}

\begin{table}[h]
\centering
\caption{Non-learned deduplication baselines versus learned heads (UCS, unseen events).}
\label{tab:baselines}
\renewcommand{\arraystretch}{1.2}\small
\begin{tabular}{lccc}
\toprule
method & R@1 & R@5 & MRR \\
\midrule
Chromaprint fingerprint & 0.066 & 0.090 & 0.079 \\
log-mel spectrogram cosine & 0.236 & 0.318 & 0.279 \\
MFCC cosine & 0.229 & 0.319 & 0.275 \\
frozen CLAP & 0.611 & 0.801 & 0.696 \\
instance head & \textbf{0.800} & \textbf{0.945} & \textbf{0.865} \\
\bottomrule
\end{tabular}
\fnote{Exact-twin retrieval on the identical full-gallery task ($N=3{,}065$, chance R@1 $=0.0003$).
Chromaprint is the standard audio-fingerprint duplicate detector; log-mel and MFCC cosine are
non-learned spectrogram baselines. All fail far below frozen CLAP and the instance head, so the
correspondence lives in learned event structure, not fingerprintable waveform overlap.}
\end{table}

\end{document}